\documentclass[twocolumn]{revtex4}
\usepackage{amsmath}
\usepackage{graphicx}

\begin{document}

\title{Entropy-variation with respect to the resistance in quantized RLC
circuit derived by generalized Hellmann-Feynman theorem}
\author{{\small Hong-yi Fan}$^{1}${\small , Xue-xiang Xu}$^{1,2}${\small and
Li-yun Hu}$^{1,2\text{*}}$}
\affiliation{$^{1}${\small Department of Physics, Shanghai Jiao Tong University, Shanghai
200030, China}\\
$^{2}${\small College of Physics \& Communication Electronics, Jiangxi
Normal University, Nanchang 330022, China}\\
Emails: hlyun@sjtu.edu.cn; hlyun2008@126.com.}

\begin{abstract}
By virtue of the generalized Hellmann-Feynman theorem for ensemble average,
we obtain internal energy and average energy consumed by the resistance $R$
in a quantized RLC electric circuit. We also calculate entropy-variation
with respect to $R$. The relation between entropy and $R$ is also derived.
By depicting figures we indeed see that the entropy increases with the
increment of $R$.

\end{abstract}

\maketitle

\section{Introduction}

In the field of mesoscopic physics Louisell was the first who quantized a
mesoscopic L-C (inductance $L$ and capacitance $C$) circuit as a quantum
harmonic oscillator \cite{1}. He made it by quantizing electric charge as
the coordinate operator $q,$ while quantizing electric current $I$
multiplied by $L$ as the momentum operator $p$. Louisell's work has become
more and more popular because mesoscopic L-C circuits may have wide
applications in quantum computer. However, Louisell only calculated quantum
fluctuation of L-C circuit at zero-temperature. In Ref.\cite{2} Fan and
Liang pointed out that since electric current generates Joule thermal
effect, one should take thermo effect into account, thus every physical
observable should be evaluated in the context of ensemble average. Besides,
since entropy increases with the generation of Joule heat, one should
consider how the resistance $R$ in RLC electric circuit affects the
variation of entropy. We shall use the generalized Hellmann-Feynmam theorem
(GHFT) for ensemble average to discuss this topic. The usual
Hellmann-Feynman (H-F) theorem states \cite{3,4}
\begin{equation}
\frac{\partial E_{n}}{\partial \chi}=\left \langle \psi_{n}\right \vert
\frac{\partial H}{\partial \chi}\left \vert \psi_{n}\right \rangle ,
\label{1}
\end{equation}
where $H$ (a Hamiltonian which involves parameter $\chi$) possesses the
eigenvector $\left \vert \psi_{n}\right \rangle ,$ $H\left \vert \psi
_{n}\right \rangle =E_{n}\left \vert \psi_{n}\right \rangle .$ For many
troublesome problems in searching for energy level in quantum mechanics,
people can resort to the H-F theorem to make the analytical calculation.
However, this formula is only available for the pure state, quantum
statistical mechanics is the study of statistical ensembles of quantum
mechanics. A statistical ensemble is described by a density matrix $\rho$,
which is a non-negative, self-adjoint, trace-class operator of trace 1 on
the Hilbert space describing the quantum system. Extending Eq. (\ref{1}) to
the ensemble average case is necessary and has been done in Refs \cite{5,6,7}%
.

Our paper is arranged as follows: In Sec. 2 we briefly introduce the GHFT
for ensemble average $\left \langle H\left( \chi \right) \right \rangle
_{e}, $ where the subscript $e\ $denotes ensemble average. In Sec. 3 based
on von Neuman's quantum entropy definition $S=-k\mathtt{tr}\left( \rho \ln
\rho \right) $ and using the GHFT we derive the entropy-variation formula
for $\frac{\partial S}{\partial \chi}$ and its relation to $\frac{\partial }{%
\partial \chi}\left \langle H\left( \chi \right) \right \rangle _{e}.$ In
Sec. 4 we use the GHFT to calculate internal energy of the quantized RLC
circuit and its fluctuation, as well as the average energy consumed by
resistance $R$. In Sec. 5 we employ the GHFT to find the relation between
entropy and $R$. By depicting figures we indeed see that the entropy
increases with the increment of resistance $R.$

\section{Brief review of the generalized Hellmann-Feynman theorem}

For the mixed states in thermal equilibrium described by density operators
\begin{equation}
\rho =\frac{1}{Z}e^{-\beta H},\text{ \ \ }\beta =\left( kT\right) ^{-1},
\label{2}
\end{equation}%
where $Z=$tr$\left( e^{-\beta H}\right) $ is the partition function ($k$ is
Boltzmann constant and $T$\ is temperature), we have proposed the GHFT \cite%
{5}. Thus the ensemble average of the Hamiltonian $H$ (which is dependent of
parameter $\chi $) is
\begin{equation}
\left\langle H\left( \chi \right) \right\rangle _{e}=\text{tr}\left[ \rho
H\left( \chi \right) \right] =\frac{1}{Z\left( \chi \right) }%
\sum_{j}e^{-\beta E_{j}\left( \chi \right) }E_{j}\left( \chi \right) \equiv
\bar{E}\left( \chi \right) ,  \label{3}
\end{equation}%
and $\left\langle A\right\rangle _{e}\equiv \mathtt{tr}\left( \rho e^{-\beta
H}\right) $ for arbitrary operator $A$ of system. Performing the partial
differentiation with respect to $\chi ,$ we have \cite{5}%
\begin{eqnarray}
\frac{\partial \left\langle H\right\rangle _{e}}{\partial \chi } &=&\frac{1}{%
Z\left( \chi \right) }\left\{ \sum_{j}e^{-\beta E_{j}\left( \chi \right)
}\right.  \notag \\
&&\left. \times \left[ -\beta E_{j}\left( \chi \right) +\beta \left\langle
H\right\rangle _{e}+1\right] \frac{\partial E_{j}\left( \chi \right) }{%
\partial \chi }\right\} .  \label{4}
\end{eqnarray}%
Then using Eq.(\ref{1}) we can further write Eq.(\ref{4}) as%
\begin{equation}
\frac{\partial }{\partial \chi }\left\langle H\right\rangle
_{e}=\left\langle \left( 1+\beta \left\langle H\right\rangle _{e}-\beta
H\right) \frac{\partial H}{\partial \chi }\right\rangle _{e}.  \label{5}
\end{equation}%
Noting the relation%
\begin{equation}
\left\langle H\frac{\partial H}{\partial \chi }\right\rangle _{e}=-\frac{%
\partial }{\partial \beta }\left\langle \frac{\partial H}{\partial \chi }%
\right\rangle _{e}+\left\langle \frac{\partial H}{\partial \chi }%
\right\rangle _{e}\left\langle H\right\rangle _{e},  \label{6}
\end{equation}%
when $H$ is independent of $\beta ,$ we can reform Eq. (\ref{5}) as
\begin{equation}
\frac{\partial }{\partial \chi }\left\langle H\right\rangle _{e}=\frac{%
\partial }{\partial \beta }\left[ \beta \left\langle \frac{\partial H}{%
\partial \chi }\right\rangle _{e}\right] =\left( 1+\beta \frac{\partial }{%
\partial \beta }\right) \left\langle \frac{\partial H}{\partial \chi }%
\right\rangle _{e}.  \label{7}
\end{equation}%
The integration of Eq.(\ref{7}) yields two forms. One is

\begin{equation}
\beta \left \langle \frac{\partial H\left( \chi \right) }{\partial \chi }%
\right \rangle _{e}=\int d\beta \frac{\partial}{\partial \chi}\left \langle
H\right \rangle _{e}+K,  \label{8}
\end{equation}
which deals with integration over $d\beta$ and $K$ is an integration constant%
$;$ and the other is
\begin{equation}
\left \langle H\right \rangle _{e}=\int_{0}^{\chi}\left( 1+\beta \frac{%
\partial }{\partial \beta}\right) \left \langle \frac{\partial H}{\partial
\chi }\right \rangle _{e}d\chi+\left \langle H\left( 0\right) \right \rangle
_{e},  \label{9}
\end{equation}
which tackles integration over $d\chi.$ Note that the fluctuation of $H$ can
be obtain by virtue of
\begin{equation}
\left( \Delta H\right) ^{2}=\left \langle H^{2}\right \rangle _{e}-\bar{E}%
^{2}=-\frac{\partial \left \langle H\right \rangle _{e}}{\partial \beta }.
\label{10}
\end{equation}

\section{Deriving entropy-variation $\frac{\partial S}{\partial \protect\chi}
$ and its relation to $\frac{\partial \left \langle H\right \rangle _{e}}{%
\partial \protect\chi}$ from $S=-k\mathtt{tr}\left( \protect\rho \ln \protect%
\rho \right) $}

Entropy $S$ in classical statistical mechanics is defined as
\begin{equation}
F=U-TS,  \label{11}
\end{equation}%
where $U$ is system's internal energy or the ensemble average of Hamiltonian
$\left\langle H\right\rangle _{e}$, and $F$ is Helmholtz free energy,%
\begin{equation}
F=-\frac{1}{\beta }\ln \sum_{n}e^{-\beta E_{n}}.  \label{12}
\end{equation}%
According to Eq.(\ref{12}) the entropy can not be calculated until systems'
energy level $E_{n}$ is known. In this work we consider how to derive
entropy without knowning $E_{n}$ in advance, i.e., we will not diagonalize
the Hamiltonian before calculating the entropy, instead, our starting point
is using entropy's quantum-mechanical definition,
\begin{equation}
S=-k\mathtt{tr}\left( \rho \ln \rho \right) .  \label{13}
\end{equation}%
It is von Neuman who extended the classical concept of entropy (put forth by
Gibbs) into the quantum domain. Note that, because the trace is actually
representation independent, Eq. (\ref{13}) assigns zero entropy to any pure
state. However, in many cases $\ln \rho $ is unknown until $\rho $ is
diagonalized, so we explore how to use the GHFT to calculate entropy of some
complicated systems, which, to our knowledge, has not been calculated in the
literature before. Rewriting Eq. (\ref{13}) as%
\begin{equation}
S=\beta k\mathtt{tr}(\rho H)+k\mathtt{tr}\left( \rho \ln Z\right) =\frac{1}{T%
}\left\langle H\right\rangle _{e}+k\ln Z,  \label{14}
\end{equation}%
where $\left\langle H\right\rangle _{e}$ corresponding to $U$ in Eq.(\ref{11}%
), it then follows%
\begin{equation}
\frac{\partial S}{\partial \chi }=\frac{1}{T}\left( \frac{\partial }{%
\partial \chi }\left\langle H\right\rangle _{e}-\left\langle \frac{\partial H%
}{\partial \chi }\right\rangle _{e}\right) ,  \label{15}
\end{equation}%
which indicates that the entropy-variation is proportional to the difference
between internal energy's variation and the ensemble average of $\frac{%
\partial H}{\partial \chi }.$ In particular, when $\rho $ is a pure state,
then $\frac{\partial }{\partial \chi }\left\langle H\right\rangle
_{e}=\left\langle \frac{\partial H}{\partial \chi }\right\rangle _{e},$ $%
\frac{\partial S}{\partial \chi }=0,$ $S$ is a constant (zero). Supposing
the case is
\begin{equation}
H=\sum_{i}\chi _{i}H_{i},\text{ }\left\langle H\right\rangle
_{e}=\sum_{i}\chi _{i}\left\langle H_{i}\right\rangle _{e},  \label{16}
\end{equation}%
then due to $\left\langle \frac{\partial H}{\partial \chi _{i}}\right\rangle
_{e}=\left\langle H_{i}\right\rangle _{e},$ we also have%
\begin{equation}
\frac{\partial S}{\partial \chi _{i}}=0.  \label{17}
\end{equation}%
Eq. (\ref{15}) also appears in Ref. \cite{6}, but it does not mention the
von Neuman entropy $S=$ $-k\mathtt{tr}\left( \rho \ln \rho \right) $.
Substituting Eq. (\ref{15}) into Eq. (\ref{7}) yields
\begin{equation}
T\frac{\partial S}{\partial \chi }=\beta \frac{\partial }{\partial \beta }%
\left\langle \frac{\partial H}{\partial \chi }\right\rangle _{e},  \label{18}
\end{equation}%
this is another form of the entropy-variation formula. It then follows
\begin{equation}
TS=\left\langle H\right\rangle _{e}-\int \left\langle \frac{\partial H}{%
\partial \chi }\right\rangle _{e}d\chi +C,  \label{19}
\end{equation}%
where $C$ is an integration constant of parameters involved in $H$ other
than $\chi .$

\section{Internal energy and average energy consumed by resistance in the
RLC circuit}

In terms of $q-p$ quantum variables $\left( \left[ q,p\right] =i\hbar
\right) $, Louisell's Hamiltonian for the quantized RLC\ circuit is
\begin{equation}
H=\frac{1}{2L}p^{2}+\frac{1}{2C}q^{2}+\frac{R}{2L}(pq+qp).  \label{20}
\end{equation}%
We now use GHFT to calculate the internal energy $\left\langle
H\right\rangle _{e}$. Substituting Eq.(\ref{20}) into Eq.(\ref{5}) and
letting $\chi $ be $L$, $C$, and $R,$ respectively, we obtain%
\begin{align}
-2L^{2}\frac{\partial \left\langle H\right\rangle _{e}}{\partial L}&
=\left\langle \left( 1+\beta \left\langle H\right\rangle _{e}-\beta H\right)
\left( p^{2}+R(pq+qp)\right) \right\rangle _{e},  \label{21} \\
-2C^{2}\frac{\partial \left\langle H\right\rangle _{e}}{\partial C}&
=\left\langle \left( 1+\beta \left\langle H\right\rangle _{e}-\beta H\right)
\left( q^{2}\right) \right\rangle _{e},  \label{22} \\
2L\frac{\partial \left\langle H\right\rangle _{e}}{\partial R}&
=\left\langle \left( 1+\beta \left\langle H\right\rangle _{e}-\beta H\right)
\left( pq+qp)\right) \right\rangle _{e}.  \label{23}
\end{align}%
Supposing the eigenvector of Hamiltonian is $\left\vert \Psi
_{n}\right\rangle $, $H\left\vert \Psi _{n}\right\rangle =E_{n}\left\vert
\Psi _{n}\right\rangle ,$ $E_{n}$ is the energy eigenvalue, due to%
\begin{equation}
\left\langle \Psi _{n}\right\vert \left[ q^{2}-p^{2},H\right] \left\vert
\Psi _{n}\right\rangle =0,  \label{24}
\end{equation}%
and%
\begin{equation}
\left[ q^{2}-p^{2},H\right] =\left( \frac{i}{L}+\frac{i}{C}\right) (pq+qp)+2i%
\frac{R}{L}\left( p^{2}+q^{2}\right) ,  \label{25}
\end{equation}%
which leads to the following relation
\begin{equation}
\left\langle \Psi _{n}\right\vert \left[ \left( \frac{i}{L}+\frac{i}{C}%
\right) (pq+qp)+2i\frac{R}{L}\left( p^{2}+q^{2}\right) \right] \left\vert
\Psi _{n}\right\rangle =0,  \label{26}
\end{equation}%
and noticing $\left\langle \beta \left\langle H\right\rangle _{e}-\beta
H\right\rangle _{e}=0,$ thus we can have the ensemble average%
\begin{eqnarray}
&&\left\langle \left( 1+\beta \left\langle H\right\rangle _{e}-\beta
H\right) \right.  \notag \\
&&\left. \times \left[ \left( \frac{i}{L}+\frac{i}{C}\right) (pq+qp)+2i\frac{%
R}{L}\left( p^{2}+q^{2}\right) \right] \right\rangle _{e}\left. =\right. 0.
\label{27}
\end{eqnarray}%
Substituting Eqs.(\ref{21})-(\ref{23}) into Eq.(\ref{27}), we obtain a
partial differential equation%
\begin{equation}
L^{2}\frac{\partial \left\langle H\right\rangle _{e}}{\partial L}+C^{2}\frac{%
\partial \left\langle H\right\rangle _{e}}{\partial C}+\left( LR-\frac{L^{2}%
}{2RC}-\frac{L}{2R}\right) \frac{\partial \left\langle H\right\rangle _{e}}{%
\partial R}=0,  \label{28}
\end{equation}%
which can be solved by virtue of the method of characteristics \cite{8,9}.
According to it we have the equation%
\begin{equation}
\frac{dL}{L^{2}}=\frac{dC}{C^{2}}=\frac{dR}{\left( LR-\frac{L^{2}}{2RC}-%
\frac{L}{2R}\right) },  \label{29}
\end{equation}%
it then follows that
\begin{equation}
\frac{1}{L}-\frac{1}{C}=c_{1},\text{ }\frac{R^{2}}{L^{2}}-\frac{1}{LC}=c_{2},
\label{30}
\end{equation}%
where $c_{1}$ and $c_{2}$ are two arbitrary constants. We can now appy the
method above in which the general solution of the partial differenial
equation (\ref{28}) is found by writing $\left\langle H\right\rangle _{e}=f%
\left[ c_{1},c_{2}\right] ,$ i.e.,

\begin{equation}
\left \langle H\right \rangle _{e}=f\left[ \frac{1}{L}-\frac{1}{C},\frac{%
R^{2}}{L^{2}}-\frac{1}{LC}\right] ,  \label{31}
\end{equation}
where $f\left[ x,y\right] $ is some function of $x,y$. In order to determine
the form of this function, we examine the special case when $R=0$, i.e
\begin{equation}
H_{0}=\frac{1}{2L}p^{2}+\frac{1}{2C}q^{2}=\hbar \omega_{0}\left( a^{\dagger
}a+\frac{1}{2}\right) ,  \label{32}
\end{equation}
where $a=\sqrt{\frac{L\omega_{0}}{2\hbar}}q+i\sqrt{\frac{1}{2\hbar L\omega
_{0}}}p$ with $\omega_{0}=1/\sqrt{LC}$. According to the well-know Bose
statistics formula $\left \langle H_{0}\right \rangle _{e}=\frac{\hbar
\omega _{0}}{2}\coth \frac{\hbar \omega_{0}\beta}{2}$, then we know%
\begin{equation}
\left \langle H\left \vert _{R=0}\right. \right \rangle _{e}=f\left[ \frac {1%
}{L}-\frac{1}{C},-\frac{1}{LC}\right] =\frac{\hbar \omega_{0}}{2}\coth \frac{%
\hbar \omega_{0}\beta}{2}.  \label{33}
\end{equation}
To determine the form of function $f\left[ x,y\right] ,$ we let $x=\frac {1}{%
L}-\frac{1}{C},$ $y=\frac{-1}{LC},$ then its reverse relations are $L=\frac{%
x+\sqrt{x^{2}-4y}}{2y},$ $C=\frac{-x+\sqrt{x^{2}-4y}}{2y},$ and $\omega_{0}=%
\sqrt{-y}$. This implies that the form of function $f\left[ x,y\right] $ is
\begin{equation}
f\left[ x,y\right] =\frac{\hbar \sqrt{-y}}{2}\coth \frac{\hbar \beta \sqrt{-y%
}}{2},  \label{34}
\end{equation}
and we obtain the internal energy

\begin{equation}
\left\langle H\right\rangle _{e}=f\left[ \frac{1}{L}-\frac{1}{C},\frac{R^{2}%
}{L^{2}}-\frac{1}{LC}\right] =\frac{\hbar \omega }{2}\coth \frac{\hbar
\omega \beta }{2},  \label{35}
\end{equation}%
where $\omega =\sqrt{\frac{1}{LC}-\frac{R^{2}}{L^{2}}}.$

Then according to Eq.(\ref{10}) the fluctuation of $H$ is
\begin{equation}
\left( \Delta H\right) ^{2}=\frac{\hbar^{2}\omega^{2}}{4}\frac{1}{\sinh ^{2}%
\frac{\beta \hbar \omega}{2}}.  \label{37}
\end{equation}
Using Eq.(\ref{8}) and the following integration formula,%
\begin{equation}
\int \frac{1}{e^{ax}-1}dx\allowbreak=\frac{1}{a}\left( \ln \left(
e^{ax}-1\right) -ax\right) ,  \label{38}
\end{equation}
we have

\begin{align}
\left \langle \frac{\partial H}{\partial R}\right \rangle _{e} & =\frac{1}{2L%
}\left \langle \left( pq+qp\right) \right \rangle _{e}  \notag \\
& =\frac{1}{\beta}\int \frac{\partial \left \langle H\right \rangle _{e}}{%
\partial R}d\beta=-\frac{\hbar \allowbreak R}{2\omega L^{2}}\coth \frac {%
\hbar \omega \beta}{2},  \label{39}
\end{align}
so the average energy consumed by the resistance is
\begin{equation}
\frac{R}{2L}\left \langle \left( pq+qp\right) \right \rangle _{e}=-\frac{%
\hbar R^{2}}{2\omega L^{2}}\coth \frac{\hbar \omega \beta}{2},\text{ }%
\omega=\omega _{0}\sqrt{1-R^{2}C/L},  \label{40}
\end{equation}
where the minus sign implies that the resistance is a kind of energy
consuming element.

\section{Entropy-variation with respect to the resistance}

In this section, based on the above results we investigate the influence of
the resistance on the entropy of RLC electric circuit. By substituting Eqs.(%
\ref{31}) and (\ref{39}) into Eq.(\ref{15}), it is easily obtained that

\begin{equation}
\frac{\partial S}{\partial R}=\allowbreak \frac{\beta R\hbar^{2}}{TL^{2}}%
\frac{\exp \left( \hbar \beta \omega \right) }{\left( \exp \left( \hbar
\beta \omega \right) -1\right) ^{2}}=\frac{\beta R\hbar^{2}}{4TL^{2}}\frac {1%
}{\sinh^{2}\frac{\beta \hbar \omega}{2}}.  \label{41}
\end{equation}
Further, making use of the integral formula

\begin{equation}
\int \frac{\ln y}{\left( y-1\right) ^{2}}dy=\ln \left( y-1\right) -\frac{%
y\ln y}{y-1},  \label{42}
\end{equation}%
we derive the relation between the entropy and the resistance as follows%
\begin{equation}
S=-k\ln \left[ \exp \left( \hbar \beta \omega \right) -1\right] +\frac{1}{T}%
\frac{\hbar \omega \exp \left( \hbar \beta \omega \right) }{\exp \left(
\hbar \beta \omega \right) -1}.  \label{43}
\end{equation}%
Obviously when $R=0$, the entropy in (\ref{43}) corresponds to LC electric
circuit. Based on the relation in Eq.(\ref{43}), we in Figure 1 depict the
variation of the entropy as a function of resistance in the range of $\left[
0,\sqrt{L/C}\right] $. The figure illustrates that the entropy has a
monotonically increasing with the resistance $R.$ When $R$ goes to the limit$%
\sqrt{L/C},$ the entropy tends to infinity.

In summary, by virtue of the generalized Hellmann-Feynman theorem for
ensemble average, we have obtained internal energy and average energy
consumed by the resistance, we have also calculated entropy-variation with
respect to the resistance in quantized RLC electric circuit. The relation
between entropy and resistance is also derived. By depicting figure we
indeed see that the entropy increases with the increment of $R.$

\textbf{ACKNOWLEDGMENTS }

The work was supported by the National Natural Science Foundation of China
under Grant Nos.10775097 and 10874174.

\newpage

\begin{figure}[tbp]
\label{Fig} \centering\includegraphics[width=8cm]{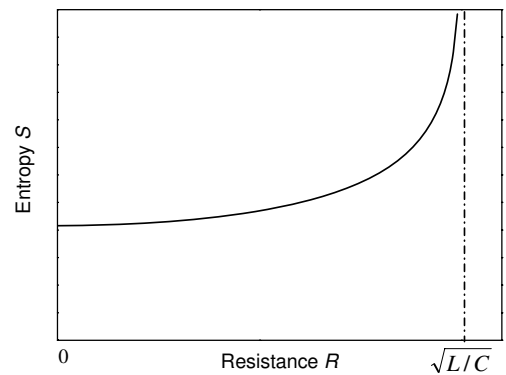}
\caption{Entropy\ S as a function of resistance.}
\end{figure}

\end{document}